\documentclass[letterpaper]{article} 
\usepackage[]{aaai2026}  
\usepackage[dvipsnames,svgnames]{xcolor}
\usepackage{times}  
\usepackage{helvet}  
\usepackage{courier}  
\usepackage[hyphens]{url}  
\usepackage{graphicx} 
\usepackage{natbib}  
\usepackage{caption} 
\usepackage{comment}

\usepackage{multicol}
\usepackage{multirow}

\usepackage{algorithm}
\usepackage{algorithmic}
\usepackage{changebar}

\newcommand{\arxivonly}[1]{#1}

\usepackage[most]{tcolorbox}
\definecolor{lightBlue}{HTML}{56B4E9}
\definecolor{darkBlue}{HTML}{1A5276}

\newtcolorbox{rqbox}[1][IMPORTANT]{
  enhanced,
  breakable,  
  colframe=lightBlue,                  
  colback=lightBlue!5,                
  coltitle=lightBlue,                 
  fonttitle=\bfseries\sffamily,       
  title=Open Research Problems,           
  detach title,                       
  before upper={\tcbtitle\hspace{0.5em}}, 
  arc=0pt,                            
  outer arc=0pt,
  boxrule=0pt,                        
  leftrule=5pt,                       
  left=4pt,                          
  right=4pt,                         
  top=2pt,                           
  bottom=2pt                         
}

\newtcolorbox{reqbox}[1][IMPORTANT]{
  enhanced,
  breakable, 
  colframe=darkBlue,                  
  colback=darkBlue!5,                
  coltitle=darkBlue,                 
  fonttitle=\bfseries\sffamily,       
  title=Elements to be documented,           
  detach title,                       
  before upper={\tcbtitle\hspace{0.5em}}, 
  arc=0pt,                            
  outer arc=0pt,
  boxrule=0pt,                        
  leftrule=5pt,                       
  left=4pt,                          
  right=4pt,                         
  top=2pt,                           
  bottom=2pt                         
}

\newcommand\blfootnote[1]{%
  \begingroup
  \renewcommand\thefootnote{}\footnote{#1}%
  \addtocounter{footnote}{-1}%
  \endgroup
}
\usepackage{newfloat}
\usepackage{listings}
\DeclareCaptionStyle{ruled}{labelfont=normalfont,labelsep=colon,strut=off} 
\floatstyle{ruled}
\newfloat{listing}{tb}{lst}{}
\floatname{listing}{Listing}
\usepackage{booktabs}

\title{Beyond Predictable Paths: AI Security Incident Reporting for Compromised Agents}
\author {
    Anastasia Pustozerova\textsuperscript{\rm 1}\equalcontrib,
    Eugene Bagdasarian\textsuperscript{\rm 2},
    Luca Beurer-Kellner\textsuperscript{\rm 3},
    Battista Biggio\textsuperscript{\rm 4},
Nico Ebert\textsuperscript{\rm 5},
David Filip\textsuperscript{\rm 6},
Marc Fischer\textsuperscript{\rm 3},
Heather Frase\textsuperscript{\rm 8,9},
David Hofer\textsuperscript{\rm 3},
Juliane Hoffmann\textsuperscript{\rm 7},
Daphne Ippolito\textsuperscript{\rm 10},
Somesh Jha\textsuperscript{\rm 11},
Sean McGregor\textsuperscript{\rm 9}, 
Esfandiar Mohammadi\textsuperscript{\rm 12},
Luca Nannini\textsuperscript{\rm 13},
Cristina Nita-Rotaru\textsuperscript{\rm 14},
Alina Oprea\textsuperscript{\rm 14},
Kevin Paeth\textsuperscript{\rm 15},
Andrew Paverd\textsuperscript{\rm 16},
Jonathan Petit\textsuperscript{\rm 17},
Andreas Rauber\textsuperscript{\rm 18},
Christian Riess\textsuperscript{\rm 7},
John Sotiropoulos\textsuperscript{\rm 19},
Andreas Wespi\textsuperscript{\rm 20},
Kathrin Grosse\textsuperscript{\rm 21}\equalcontrib
}
\affiliations {
    \textsuperscript{\rm 1}SBA Research, Austria 
    \textsuperscript{\rm 2}University of Massachusetts Amherst, US
    \textsuperscript{\rm 3}Snyk, Switzerland 
    \textsuperscript{\rm 4}Universita degli studi di Cagliari, Italy
    \textsuperscript{\rm 5}ZHAW, Switzerland 
    \textsuperscript{\rm 6}ISO/IEC JTC 1/SC 42, Huawei, Ireland 
    \textsuperscript{\rm 7}FAU Erlangen-Nürnberg, Germany 
    \textsuperscript{\rm 8}Veraitech and Virginia Tech, US 
    \textsuperscript{\rm 9}Responsible AI collaborative, US     
    \textsuperscript{\rm 10}Carnegie Mellon University, US
    \textsuperscript{\rm 11}University of Wisconsin, US 
    \textsuperscript{\rm 12}University of Lübeck, Germany 
    \textsuperscript{\rm 13}Trustora Digital, Spain
    \textsuperscript{\rm 14}Northeastern University, US 
    \textsuperscript{\rm 15}UL Research Institutes, US 
    \textsuperscript{\rm 16}Microsoft Security Response Center (MSRC), Great Britain 
    \textsuperscript{\rm 17}Qualcomm, US 
    \textsuperscript{\rm 18}TU Vienna, Austria 
    \textsuperscript{\rm 19}Deep Cyber/OWASP GenAI Security Project, United Kingdom
    \textsuperscript{\rm 20}IBM Research Europe--Zurich, Switzerland
    \textsuperscript{\rm 21}Independent, Germany\\  
    kathrgrosse@gmail.com
}

\usepackage{bibentry}

\begin{document}

\maketitle

\begin{abstract}
AI agents are being deployed rapidly, accompanied by a growing number of AI-specific attacks and corresponding incidents. As incident reporting becomes increasingly important for legal compliance, governance, accountability, and security; current frameworks must be adapted to the unique characteristics of AI agents. In this paper, two editorial authors compare AI systems and AI agents and, drawing on input from \textbf{23 experts} in academia and industry, identify the information required for reporting incidents where the security of  AI agents is harmed.  
Potential reporting elements include, for example, agent memory and memory accesses, actual and potential levels of autonomy, and tool usage.
Based on these findings, we identify several open research questions, including how to efficiently record incidents and how to determine whether vulnerabilities and incidents generalize. Expert feedback also highlighted potential reporting weaknesses, such as risks of data leakage and attacks targeting the reporting infrastructure itself, creating additional research needs. Lastly, we summarize privacy requirements and outline research directions for the secure and trustworthy deployment of AI agents.
\end{abstract}

\blfootnote{This paper synthesizes diverse expert perspectives. Contributors provided input within their specific domains of expertise; authorship does not imply endorsement of every sub-section.}

\section{Introduction}
\begin{figure}[ht]
    \centering \includegraphics[width=0.95\columnwidth]{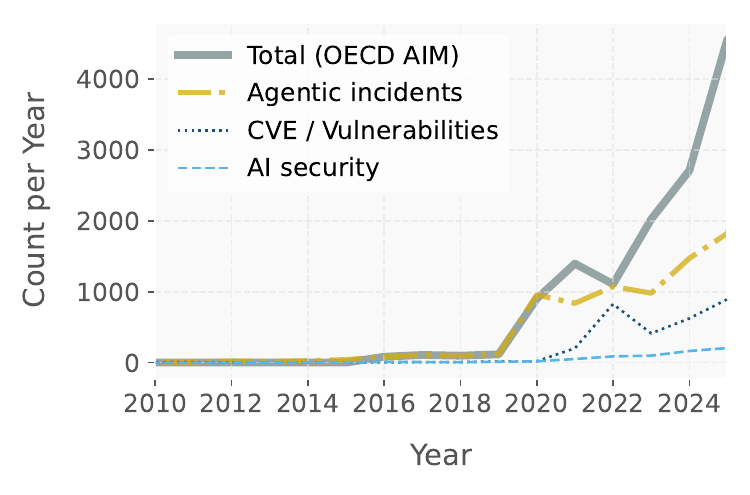}
    \caption{\textbf{AI Incidents and vulnerabilities}. We plot the total number of incidents in the OECD AI incident monitor (AIM) (gray)\arxivonly{~\cite{oecdaim}}, which tracks and documents real world AI incidents and hazards. Agentic incidents (yellow, dash-doted) are shown separately following    \citet{guilherme_jr_2026_20248676}. We further distinguish incidents with an officially assigned vulnerability (dark blue, dotted) or an AI security tag (light blue, dashed), also based on \citet{guilherme_jr_2026_20248676}.}
    \label{fig:incidents}
\end{figure}

AI agents are increasingly targeted by attackers through prompt injection, memory poisoning, tool compromise, and other agent-specific attack vectors.
This is reflected in an increasing number of AI incidents, which also entail non-security related incidents involving agents.
As visible in Figure~\ref{fig:incidents}, a large fraction of all AI incidents involve large language models (LLMs) and agentic AI (yellow). A significant fraction of these are security-related, and intentionally exploit known vulnerabilities (CVE, dark blue) or tamper with the AI models or agents specifically (light blue).

Security vulnerabilities for AI agents span the agent's data (EchoLeak, CVE-2025-32711; ShareLeak in Copilot Studio, CVE-
2026-21520), context (Reprompt, CVE-2026-24307), and a accessed tools (GitHub Copilot, CVE-2025-53773). 
Scientific work further confirms
systematically misdiagnosed risks at, for example, the sequence of steps the agent takes\arxivonly{~\cite{li2026atbench}}.

These findings raise the question of how such incidents and vulnerabilities can guide incident response and inform security best practices.
For example, research in non-agentic AI security has played a 
critical role in improving AI robustness, uncovering hidden model vulnerabilities, and driving more reliable evaluation methodologies. Concepts such as threat modeling and adaptive attacks are now foundational to AI red teaming and security engineering.
 To systematically learn from real-world incidents, an appropriate structured form of incident reporting needs to be established. Such reporting is 
 indeed already demanded by legislation like the AI Act Article 73 reporting obligations\arxivonly{~\cite{EUAIAct2024}}. 
 AI agents, as discussed in detail below, constitute a specific subset of AI systems characterized by multi-step incidents involving multiple components and dynamic interactions. They are not adequately captured by existing approaches of incident reporting.

Fully describing an AI agent routinely includes information like functions and permissions, and may further contain (state of) memory, context length, connections to other agents,
tools, skills, and capabilities, including their versions, accessible data and APIs, code, supply-chain dependencies, and as well as any messages exchanged,  intermediate states of sub-agents, logs, and tool traces. 
Because agentic AI systems dynamically compose such tools, sub-agents, APIs, and external data sources at runtime, the conditions giving rise to an incident may never have existed during development or testing. Incident reporting therefore becomes a foundational mechanism for uncovering new risks and preventing future incidents\arxivonly{~\cite{wei_designing_2026,gailmard_known_2025}}.

While there are proposals for non-agentic AI security incident reports \arxivonly{~\cite{bieringer,strom2018mitreattack,fazelnia2024establishing}}, or reporting for agentic AI for non-security incidents~\cite{ezell2025incident}; there is no security reporting standard for AI agents. Our contributions are thus \arxivonly{six-fold}:
\begin{enumerate}
    \item We show, based on definitions of the OECD and the EU, that the difference between AI systems and AI agents is small but substantial for security incident reporting.
    \item We consolidate the perspectives of \textbf{\arxivonly{23}} experts on \arxivonly{concerns and challenges in} agentic AI security incident reporting.
    \item We derive what information needs to be collected for AI agent security incidents (see Table~\ref{tab:overviewElements}).
    \item We identify open research questions related to security incident reporting for AI agents.
    \item We highlight novel security and privacy challenges that apply to agentic AI security incident reporting. 
    \arxivonly{\item We summarize implications for research, legislation, and standards.}
\end{enumerate}

\arxivonly{This paper is organized as follows.  
We first review necessary background, defining AI systems and agents, agentic security, incident, and summarizing related work in research, standards, and legislation. We then discuss how to describe an agentic security incident. After presenting specific security and privacy properties needed for security incident reporting, we outline implications for both research and legislation and standards and conclude. }
\section{Background}
Before discussing incident reporting for AI agents, we define what we mean by an AI agent, what security risks exist for such systems, how to define the corresponding incident, and what related scientific and standards work exist.

\subsection{Defining AI systems and AI agents}
The OECD definition of an AI system\arxivonly{~\cite{OECDAiSystemDefMemo2024}} provides the conceptual foundation for the EU AI Act and work on AI incident reporting~\cite{bieringer}.
The OECD defines ``\emph{an \textbf{AI system} [as] a machine-based system that, for explicit or implicit objectives, infers, from the input it receives, how to generate outputs such as predictions, content, recommendations, or decisions that can influence physical or virtual environments. Different AI systems vary in their levels of autonomy and adaptiveness after deployment.}''

In this work, the term AI agent refers to the modern agents based on large language models (LLM)\arxivonly{~\cite{sapkota2025ai}}, which are a subset of AI systems. \textbf{AI agents} are  defined by the European Union as
``\emph{agentic artificial intelligence (agentic AI) [is] a concept in artificial intelligence (AI) that describes systems acting autonomously with limited human interactions (in particular, without step-by-step instructions) to fulfill goals rather than isolated tasks}.''\footnote{\url{https://www.edps.europa.eu/data-protection/technology-monitoring/techsonar/agentic-ai_en}} 

In other words, AI agents (or agentic AI) are a specific subset of AI systems. They shift from task-driven execution to goal-directed behavior. 
According to Merriam Webster, a task is ``\emph{a piece of work that has been given to someone}'', whereas a goal is ``\emph{the end toward which effort is directed, [an] aim}''. 
A goal thus requires many steps that the agents carries out. In contrast, non-agentic AI systems focus on tasks, often solvable in a single or few interactions.

This high-level difference between AI systems and AI agents mandates changes in how to describe an event involving such an agent (i.e., an incident, as defined later).
Unlike traditional non-agentic AI systems, where an event possibly relies on a single input-output pair, an agentic incident unfolds across a sequence of reasoning steps, tool invocations, memory accesses, and environmental interactions. 
Hence, agent-based interactions have a strong \textbf{trajectory}-level  
character~\cite{bisconti2026boiling}.
A second distinguishing property 
is the expanded \textbf{capabilities} of an AI agent required to achieve goals rather than isolated tasks. 
Agents can cause direct effects on their environment, for example, on the underlying operating system and connected services. Thus, their potential impact is significantly broader.
In addition, despite AI agents being a subset of AI systems, they exhibit structural differences, including the ability to comprise of \textbf{multiple AI models} or even other AI systems\arxivonly{~\cite{li2023camel}}.

While none of these differences is strictly unique to agentic AI, each is considerably more pronounced in agents than in non-agentic AI systems. 

\subsection{Attacks on AI agents}
Following ISO/IEC 27000:2018\arxivonly{~\cite{isosecruityRef}}, we define an attack as an attempt to destroy, expose, alter, disable, steal an asset, or to gain unauthorized access to or make unauthorized use it. Such attacks can target any computer or digital system. An underlying flaw or weakness that can be exploited in an attack is often formally cataloged and assigned a unique Common Vulnerabilities and Exposures (CVE) identifier to facilitate standardized tracking and remediation. AI systems suffer from such flaws as any other software system. In addition, AI systems are vulnerable to a plethora of AI specific attacks, including training or test data tampering\arxivonly{~\cite{biggio2018wild}}, and the introduction of backdoors or bias\arxivonly{~\cite{cina2023wild}},
sensitive data inference\arxivonly{~\cite{oliynyk2023know}}, or sloth attacks\arxivonly{~\cite{brachemi2026energy}}. 

Moreover, there are specific attacks on AI agents.
For example, external data used by an agent to solve a partial task can be \emph{poisoned} with malicious instructions, allowing an attacker to influence the agent without requiring direct control over it.
Unlike classical data poisoning, however, this occurs at inference time rather than during training. More broadly, an action may directly originate either from an initial prompt or from a reasoning process and can be significantly influenced by external data. 
Hence, seemingly innocuous prompts and data may \emph{unfold} into an attack during reasoning, i.e., the reasoning process is used to cause harm.
Agents' inherent autonomy enables attacks such as indirect prompt injection, tool misuse, memory poisoning, and multi-agent trust exploitation. Such attacks form complex exploit chains
involving privilege escalation and data exfiltration\arxivonly{~\cite{greshake2023not,reddy2025echoleak,cve-2024-5565}}. 
This expanded attack surface fundamentally alters the 
threat model applicable to classical AI systems. 

Security on AI and AI agents has so far produced extensive academic literature, yet relatively few documented operational incidents. Studies have shown that attackers overwhelmingly prefer cheaper and more reliable methods — such as phishing, credential theft, and supply-chain compromise — over sophisticated attacks optimized against the target model, which have not yet been weaponized on a large scale\arxivonly{~\cite{apruzzese2023real,grosse2024your}.}
Despite their currently low prevalence, such attacks warrant proactive mitigation due to their potentially high impact, which may be further amplified by the autonomy and capabilities of agentic systems.

\subsection{Defining AI Incidents}
The \arxivonly{\citet{oecdDefiningAI2024}} defines an AI incident 
as an event involving an AI system where harm incurred. Examples of harm include injury, disruptions of critical infrastructure, violations of human rights, and harm to property. Other definitions of incidents vary depending on how close an event comes to causing harm. Even "near misses" -- events that could have caused harm but did not -- can be a valuable source of learning~\cite{wei_designing_2026}.
Such harm-event based reporting is usually implicated by model failures (e.g., biased outputs, hallucinations, privacy leakage).
In contrast, reporting within non-AI security 
assumes 
a clear attacker-victim topology: a malicious actor exploits a vulnerability in a system, and the system's operator is the harmed party. 
Patching vulnerable systems typically reduces their future vulnerability. 
Hence, Reporting (AI) incidents for safety and security differs~\cite{bieringer2024position}
\arxivonly{, as safety aims to protect external actors from the AI system, whereas security protects the AI system from malicious external actors\arxivonly{\cite{qi2024airiskmanagementincorporate,Khlaaf2026TowardCR}}. Security} reporting requires more attacker-centered information~\cite{bieringer2024position}. 

In the remainder of the paper, we focus on security incidents. The agent description can potentially be used for safety incidents involving agentic AI, or for incidents where the AI caused an incident. Examples are recent incidents involving OpenAI\footnote{\url{https://openai.com/index/hugging-face-incident-and-the-road-ahead/}} and Anthropic\footnote{\url{https://www.anthropic.com/news/investigating-incidents-cybersecurity-evals}}.
In these cases, agents exceeded their intended operational boundaries and gained unauthorized access to external systems.
These developments underscore the need for systematic agent descriptions.

\subsection{Related Scientific Work} There is limited related work on incident reports for AI agents. Previous work by MITRE\footnote{\url{https://ai-incidents.mitre.org}} and \citet{bieringer} covers only AI systems. \citet{fazelnia2024establishing} propose a minimal set of reporting elements for AI vulnerabilities\arxivonly{, and \citet{cattell2024coordinated} described a  security-centered CVE-like reporting.} Other examples for data collection taxonomies and databases focusing on AI are the AI incident database\arxivonly{~\cite{mcgregor2021preventing}}, AVID, or MITRE Atlas.
In addition, \citet{ezell2025incident} propose an incident analysis framework that, however, focuses on non-security incidents. 
The current work instead focuses on security incidents and on the agent-system gap, 
which warrants substantial changes to existing approaches. Due to the novelty of the area, few works are covering the specifics of security incident reporting for AI agents. This security focus has concrete implications: in contrast to \citet{ezell2025incident}, we suggest also documenting reads and writes to memory to identify memory tampering, tracking delegation and trust boundaries to identify the origin of a breach, and finally collecting multi-agent architecture and communication to identify threats from emergent behavior.

\subsection{Related Laws and Standards}
There are standards addressing AI and incident reporting, some of which are still under development. As our findings may inform ongoing and future standardization efforts, we provide background in terms of legal requirements and related standards.

The GPAI Code of Practice Commitment 9 on serious-incident reporting\arxivonly{~\cite{EuropeanAIOffice2025}} is the  most operationally detailed EU instrument on incident reporting, though  it operationalises Article 55(1)(c) for providers of general-purpose AI  models with systemic risk and reports to the AI Office rather than through the Article 73 channel. 
Article 73\arxivonly{~\cite{EUAIAct2024}} obliges providers of high-risk AI systems to report serious incidents and applies from 2 August 2026, while the Digital Omnibus amendments defer the Chapter III high-risk requirements to 2 December 2027 (Annex III) and 2 August 2028 (Annex I).
In general, the GPAI Code acknowledges that information about an incident cannot be reconstructed only retrospectively. This aligns with our efforts to understand up-front which information should be collected.
The doctrinal architecture for continuous compliance evaluation that Article 72 of the AI Act partially anticipates is developed at greater length in adjacent recent work~\cite{nannini2026ai}. Under Article 75(1a), providers within the AI Office's exclusive competence, which covers systems built on a general-purpose AI model by the same provider, report serious incidents to the AI Office rather than to national market surveillance authorities.

Relevant key international standardization documents include \arxivonly{~\citet{isoiec8200-2024}~}8200 and \arxivonly{\citet{isoiec42105-fdis}~}42105. The former gives the basics of how to make an AI system controllable by an external agent. TS 8200 assumes AI systems and agents as defined in ISO/IEC 22989:2022 and deliberately does not presume control by a human operator. All that matters is that the control is practically possible and the controller is external to the controlled system.
Most AI systems cannot be directly controlled by human operators because critical changes in their internal state are neither observable by humans nor interpretable in real time, making timely and meaningful human intervention difficult or impossible.
The latter \arxivonly{~\citet{isoiec42105-fdis}~}42105 describes how to ensure that AI systems' operations can be overseen by humans on the operational level without relying on other ambiguous and unverifiable criteria. 
At the same time, 
new standards are being developed, 
including protocol 
standards for agentic interactions (NLIP by ECMA TC56; IETF, LF A2A). However, also existing legislation that seems superficially unrelated can become relevant. As an example, consider GDPR\arxivonly{~\cite{gdpr2016}} and its requirements on privacy, that incident reports will have to respect, too.   

To conclude, the exact interactions, content, and requirements of these emerging standards have yet to be defined.
\section{Describing Agentic Incidents}
The background section identifies three key distinctions between AI agents and conventional AI systems: trajectory-level behavior, expanded capabilities, and the use of multiple AI models. Before examining their implications, we outline the goals of incident reporting, including documentation, reproducibility, and forensics and general technical challenges, as well as human and social aspects. We then discuss how the difference between AI agents and systems shape the information in agentic AI incident reports, also summarized in Table~\ref{tab:overviewElements}.

\textbf{Reproducibility versus Forensics.} 
The precise goal of incident reporting\arxivonly{~\cite{ebert_learning_2025}} remains undefined. Beyond mere documentation,
incident reporting could serve to reproduce an incident. It may also provide information relevant to auditing and liability assessments.
Previous work has also found that
 incident reporting
can support the identification of new malicious actors\arxivonly{~\cite{wagner2019cyber}}.
Furthermore, verifiable forensics evidence may be necessary to support legal proceedings and meet applicable evidentiary requirements\arxivonly{~\cite{CodiceProceduraPenaleArt220,AdvisoryCommittee2025,Daubert1993}.} Ensuring such forensic readiness should therefore be an integral part of the agentic system lifecycle rather than an ex post facto activity.
 Such requirements have to be weighted against intellectual property and other privacy requirements, for example, those imposed by the GDPR.

\textbf{Technical Challenges.} The incident reporting covers potentially parallel-running distributed systems and their challenges, like delays, inconsistent clocks and their implications, and data inconsistencies across instances, for example.

\textbf{Sociotechnical Challenges.} Incidents occur
at the intersection between humans and agents.
There may thus be no single truth of what happened, but different `realities'~\cite{ebert2023learning}. This is further amplified by both the complexity of agents and their cultural and psychological interplay with a user. 
Another important aspect of the real-world deployment of agents 
is their potential to operate across multiple jurisdictions. 
This may require compliance frameworks that allow agents to interact only with systems that meet common legal or policy requirements. For example, an agent may need to verify that another subsystem complies with the GDPR before sharing personal data with it.
How such cross-jurisdictional issues should be addressed legally and technically remains an open question, highlighting the need to investigate non-technical aspects of agentic AI security incidents.

\begin{rqbox}
Enabling the reproducibility and forensic analysis though incident reporting and understanding trade-offs with intellectual property and privacy are areas for future research. Such research has to consider the utility of the reports for different stakeholders, including practitioners, companies, and legal entities.Sociotechnical aspects like the psychological interaction with humans should be studied, as well as a legal and technical understanding derived about agents potentially spanning several jurisdictions. 
\end{rqbox}

\subsection{Describing the agent and its trajectory}
Due to the trajectory-like nature,
the incident should be described as a trajectory that ultimately leads to the incident-relevant state.
The existence of such trajectories has been recognized in EU law, Articles 12, 19 and 26(6)\arxivonly{~\cite{EUAIAct2024}}, requiring logging throughout the system's lifetime, with logs retained for at least six months. However, these provisions do not prescribe specific fields for tool calls, memory writes, or delegation, and the six-month minimum retention period may expire before an incrementally seeded compromise is recognized.
In the interest of the reporting party and for security reasons, logging may thus be more extensive in practice than legally required.
\textbf{Runtime trace.} Any available runtime trace and state should be collected, such as message histories,
connected systems, and execution logs. These latter entail tool use and subsystems, which we discuss in detail later.
The documentation of this runtime trace must be extensive to enable traceability and attribution.
Also output tokens for reasoning need to be saved. 
Monitoring these reasoning traces and communication with the user can, for example, help identify attacks targeting the reasoning process\arxivonly{~\cite{hu2026rethinking}}. However, such extensive logging entails trade-offs between forensic value, storage and processing overhead, and data-protection requirements such as data minimization that are yet to be investigated in depth.

Even more relevant, however, is the agent's input, or context, which is defined as all tokens available for processing by the AI model within the agent\arxivonly{~\cite{vaswani2017attention}}.
Collecting as much of this context as possible, especially over the trajectory of the agent, is crucial, especially for more complex agents\arxivonly{~\citep{GuoCWCPCW024,hadfield2025multiagent}}. Possible vulnerabilities within the context
are rather indirect. 
Instead of targeting the model itself, an attack exploits weaknesses in context handling and in model output.
As manipulated context may exist only temporarily before or during execution and may not persist after the incident~\cite{ferrag2025threats},
it is crucial to store the full context in a tamper-proof format.

\textbf{Memory and external data sources.}
Although memory and external data constitute a subset of the context, we describe it separately in more detail.
In contrast to the agent's runtime trace, most of the memory, or external data sources, are static. They need to be recorded with either a versioning number, or in full if no versioning is available or the memory is writable by the agent. Also, the full provenance of any retrieved or injected external data must be recorded.
Despite an implicit entailment in the provenance, we suggest to explicitly document the data sources accessible to the agent and the associated trust management mechanisms\arxivonly{~\cite{ferrag2025threats,lin2026,das2026}.}
While the features described above are relatively static, memory resources may change over time, and these changes should be recorded in a tamper-resistant manner.

\textbf{Autonomy.}
Agentic AI autonomy influences greatly the trajectory leading to an incident. 
It increases the velocity of the agents' progress and at the same time the difficulty for human oversight and intervention. Autonomy can vary in degree and is therefore best understood as a spectrum along which an agent can be characterized.
For example, if an agent wrote code that made private data public, but a human ultimately pressed ``approve" on this code and opted to deploy it, then this is on one side of the spectrum. 
At the other side of the spectrum, a user may authorize an AI agent to act continuously without human intervention, allowing it to execute insecure code without prior human oversight.
This spectrum is defined during the design phase, describing theoretical (intended) autonomy of an agent. This autonomy may vary by action type, component, accessed tool, or point of execution. 
Consequently, there is a concrete autonomy instantiation on this spectrum that defines the agent's autonomy throughout the trajectory leading to the incident. 
This effective autonomy is crucial for understanding the incident.

\begin{reqbox}
 are the agent's context and trajectory, including runtime trace, memory, external data provenance and the agent's autonomy level by design and in effect over time.
\end{reqbox}

\begin{rqbox}
Efficiently collecting, processing, versioning, and storing the information required to describe the agent, as well as automatically detecting security incidents, remain open research questions. 
Reducing the volume of data that must be stored while preserving information in a tamper-resistant manner requires further research, although some approaches to tamper resistance have been explored~\cite{avizheh2026magiq}.
Lastly, future work has to tackle how to document autonomy.
\end{rqbox}

\subsection{Describing Agentic Capabilities}
The way that agents interface with their surrounding is also a crucial difference from non-agentic AI. While AI systems may also interact with their environment, agents' capabilities are more complex, with more tool interactions, in potentially shorter time-frames.
Here, we focus on tools, impact on the underlying operating system, and delegation. 

\textbf{Tools.} The usage of tools is a defining characteristic of AI agents and major source of security risks and implications. 
 Similar to autonomy, tool usage in agentic systems is not a fixed property established at development time but a dynamic decision made by an AI agent at runtime. 
In other words, the theoretical ability 
to use a tool and its actual use during the incident may differ. Thus, it is crucial to document actual tool use during the incident~\cite{ezell2025incident}. 
The difference between planned and actual tool use may vary greatly, as has been found, for example, in Android security, where applications often request more capabilities (e.g., App permissions) than they actually use at runtime\arxivonly{~\cite{felt2011android}}. 

Tracking the exact tool and tool use is relevant in several scenarios. 
 An incident may occur due to tool malfunction, wrong tool selection, incorrect arguments, or inappropriate tool sequence.
 Tools can serve as vectors for data poisoning or other forms of compromise, like, for example, credential-stealing malware. At the same time, trusted and benign tools may lead to an incident if the calling agent is prompt injected.
Such subtle differences in tool use have to be distinguishable to understand incidents and prevent similar events in the future.
Lastly, we need to keep track of any tool's authentication tokens to allow traceability, attribution,
identity management and permissions.

\textbf{Impact on OS and environment.}
A consequence of tool use is that an agent's impact may extend beyond a single tool to the underlying operating system and broader system environment.
In contrast to classic machine learning that outputs a prediction or generates content, an agent can interact with and affect an underlying system in a more complex way. For example, a coding agent could produce insecure code that is executed or deployed. 
Agents may therefore also contribute to non-AI security incidents.
 Several researchers have consequently argued that agentic AI security must be approached as a systems security problem: 
the AI model powering the agent must be treated as an untrusted component, and security invariants must be enforced at the system level~\cite{christodorescu2026agent}. 
The extensive body of research in operating systems, networks, formal methods, and adversarial machine learning provides a set of core principles, grounded in decades of systems security research, that form a foundation for designing agentic systems with predictable security guarantees. 

An example of how classical security is applicable to agents 
is the semantic gap. 
Consider, for example, web agents. Low-level events (e.g. user clicked on a certain region of the webpage) are often not suitable for security incident reporting, because they lack interpretable semantics~\cite{piet2026web}. Ideally, these low-level events should be mapped to high-level events (e.g. user clicked on the approve button on the webpage), which can then be meaningfully reported. Such mappings may allow established non-AI incident-reporting approaches to be applied to agent actions; however, the boundary between AI and non-AI security incidents remains difficult to define~\cite{bieringer2024position}.

\textbf{Delegation.}
In the context of tool work, autonomy takes a special twist. A task received by an agent may have been delegated by a human user or another agent.
This inadvertently raises the question of identity management and documentation.
For example, we need to document whether the agent has been operating on behalf of a specific
user (i.e., with the permissions and access of that user), or as a separate non-human identity
(i.e., with a potentially different set of permissions).
The identity used for each tool call at the time of the incident is therefore important for understanding the incident and determining whether it resulted from improper identity management or the circumvention of access controls.

A related question concerns the origin of the incident. Although the initiator of the process may have caused the incident, it could as well have been caused by the delegate's input data, a malicious delegate, a malicious tool, faulty communication across agents, or issues in the agent's deployment. Documenting delegation to allow determining these differences is thus crucial.
Lastly, delegation necessitates clear trust boundaries between user and system, since an AI agent can faithfully execute dangerous instructions from a trusted, authenticated user.
Based on the threat model considered, the latter example may or may not constitute a security incident\arxivonly{~\citep{ayzenberg2025ruleoftwo,invariant2025githubmcp,willison2025lethaltrifecta}.}
Consequently, trust boundaries between components, including how trust is managed during delegation and communication, should be documented for incidents.

\begin{reqbox}
are possible and actual (over time) tool use, interactions with the underlying operating system, original entity that initiated an action, defined trust boundaries.    
\end{reqbox}

\begin{rqbox}
Describing tool use and the agent's effect on the operating system and broader environment without falling into the ``semantic gap'' between low- and high-level events presents an interesting and challenging problem~\cite{piet2026web}, 
for which existing solutions from non-AI security may be adaptable\arxivonly{~\cite{fu2012space}}.
Similarly, tracking the cause of an incident though delegation involving authentication tokens remains an open problem and raises again the trade-offs between protecting confidential information and the need for effective incident reporting. Solutions developed for managing privacy–utility trade-offs in data release may be useful for this line of research\arxivonly{~\cite{nanayakkara2022visualizing}}.
Lastly, the interaction between the agent and the environment raises the question of when to use traditional security reporting or reporting designed for agentic AI incidents.
\end{rqbox}

\subsection{Agent subcomponents and orchestration}
The composition of an agentic system routinely includes other AI systems such as sub-agents, retrieval models, classifier guards, and judge models, each with their own functions and permissions. These subcomponents may further be characterized by their own runtime trace, state, memory, context, as stated in the previous section, and additionally connections to other agents. It is also  possible to combine the APIs of two existing distinct agents in a new agent, or generate code, tools, even an transient agents.
Fully describing such a system 
for incident reporting purposes, therefore, requires accounting for all of these components and agents, and their relationships,
as we discuss in this subsection.

\textbf{Communication across components.}
While the composition may be somewhat fixed during design, an orchestrator agent dynamically decides at runtime 
which sub-agents to invoke. Multi-agent coordination introduces distinct failure modes 
including miscoordination, conflict, and collusion\arxivonly{~\cite{hammond2025multi}}, where the latter describes a malicious, secret agreement.
 As with other distributed systems, communication and information-exchange patterns should be carefully logged or represented to enable the monitoring and reconstruction of potential failure modes\arxivonly{~\cite{de2025open}}.
This may imply that components must 
 implement certain tamper resistant reporting requirements to form part of an AI agent and its permitted interaction environment.
In addition, communication protocols and mechanisms used to  coordinate autonomous execution need to be documented\arxivonly{~\cite{li2024survey,yan2025communication}}.
More concretely, orchestration protocols effectively become part of the attack surface or enable incidents as they govern delegation, context propagation, tool invocation, and inter-agent communication\arxivonly{~\cite{louck2026protocols}}.
Concrete security examples in this context are insufficient authentication, weak message integrity guarantees, missing provenance metadata (e.g. access rights to a document), unsafe serialization, unclear authority semantics, or an ambiguous distinction between instructions and data. 
In particular, jailbreaks or malicious instructions may propagate autonomously between agents through standard communication channels\arxivonly{~\cite{lee2024promptinfection}}.
As protocol and orchestration exploits present a major emerging class of agentic AI vulnerabilities~\cite{ferrag2025threats}, incident reporting should capture relevant information about the agentic AI system organization and inter-agent interaction. 

\textbf{Emergent behaviors.}
Agentic AI incidents and vulnerabilities can be a pure run-time artifact,
emerging from the dynamic interaction between components during execution through the agent's reasoning and actions, which can be non-deterministic.
Each of the internal components or
models may perform as expected, 
while a vulnerability or an incident may arise from some interaction between these systems. In other words, no single model is vulnerable, but the vulnerability is an implicit property of the system as a whole.

\begin{reqbox}
are the defined and actual (at runtime) multi-agent architecture, inter-agent communication, associated identities and permissions, information exchange patterns, orchestration and communication protocols and protocol versions.   
\end{reqbox}

\begin{rqbox}
Understanding how to document incidents in the context of emergent behaviors and
how to enforce documentation in each component
 remains an open research question. 
This holds particularly as it is unclear how long pre- and post activity has to be logged to enable incident identification. 
\end{rqbox}
\section{Security and privacy risks of reports}
Agents, with their complexity, autonomy and understanding of natural language, bring unique challenges to AI security incident reporting. We discuss these challenges, focusing on the security of incident reports and confidential data leakage.

If programmed or compromised to do so, agents may manipulate logs or access patterns to conceal malicious activity.
In particular, an attacker 
who is aware of the reporting scheme may deliberately craft attacks that will make them invisible in incident reports. 
That poses an additional threat to the integrity and transparency of the reporting process. How to ensure the completeness and reliability of incident reports therefore remains an open research question. While attacks targeting forensic methods are well studied in non-AI security\arxivonly{~\cite{jain2014anti}}, they remain largely unexplored in the context of AI agents, 
which may alter their behavior based on report filing. 
Nevertheless, adapting established security measures from traditional systems 
is a promising research direction.

Another threat assumes an attacker may target a critical system that mandates reporting with the sole goal of causing intellectual property or confidential information to be disclosed through the resulting incident report. This highlights an important distinction between the evidence that needs to be logged for incident reconstruction and the information ultimately disclosed in an incident report. Extensive logging, which may be required to support reproducibility and intelligence gathering, can capture sensitive information that is neither necessary nor desirable to disclose. Determining what evidence should be reported and at what level of granularity therefore remains an open research question.
A solution to such a release of information is to publish reports only in aggregate. However, 
knowing such release rules enable Sybill attacks. 
In other words, carefully crafted reports could be combined with publicly available aggregate information to infer sensitive details about a competitor's incident, potentially resulting in a confidentiality breach.

A further security risk arises when information protected as intellectual property or under regulations such as the GDPR is hidden in reports using steganography\arxivonly{~\cite{agarwal2013text}}. While this is not a new threat, larger amounts of reported information facilitate information hiding\arxivonly{~\cite{ker2008square}}. As agentic AI incident reports may contain more information, they may be particularly susceptible to such attacks.

Lastly, when agents are used for incident report filing or analysis, vulnerabilities in these agents may be exploited to disclose or embed information that was not intended for release.

\begin{rqbox}
The security, confidentiality and completeness of incident reports, as well as corresponding features of the possible agent reporters, remain important areas for future research. This includes determining what evidence should be logged and preserved and what subset should ultimately be disclosed in incident reports. 
\end{rqbox}
\section{Implications}
AI (security) incident reporting is a rapidly evolving field, driven by continuous changes in technology, legislation, and deployment settings.
Agents have the potential to largely increase the breadth 
of harms, as well as the complexity of incident identification, documentation, and reporting. Managing the incurred risks is thus crucial and requires further research into appropriate incident reporting approaches tailored to agentic AI systems.
Moreover, regulators need appropriate technical standardization tooling to effectively achieve their policy goals.
We therefore discuss the implications for research,standardization and legislation.

\subsection{Open research questions}
Our work highlights significant gaps in the state of the art that need to be addressed to allow efficient and effective incident reporting for agentic AI. 

For example, it remains an open question
what information within an incident report is mandatory and what should be optional.
This is crucial in the context of other open questions, such as what information is needed to support reproducibility, forensics, and incident generalization, 
while avoiding
an administrative burden that leads to secondary failures.    
In addition, forensic methods and other approaches are required to enable efficient and automated identification of agentic incidents. 
Beyond this technical details, incidents occur in
socio-technical systems at the intersection between humans and agents.
There may thus not be a single truth of what happened, but different `realities'~\cite{ebert2023learning}. This is further amplified by both the complexity of agents and their cultural and psychological interplay with the user~\cite{liu2024understanding}.
Another dimension where the embedding in the real world
becomes obvious
is that agents may span jurisdictions. This may imply the need of a compliance framework, allowing only agents to participate if they all follow some common guidelines, obey certain policies, or protocols. An example could be agents having to verify GDPR compliance of all other systems they share information with. 
How such problems should be handled legally and can be handled technically is still open. In summary, also non-technical aspects should be researched.

Lastly and most importantly, we should understand emerging vulnerabilities in the incidents reports themselves, like data leakage or the exploitation of reporting structures to design attacks that evade or undermine incident reporting for agentic AI systems. Assessing such threats in general but also with respect to a concrete report scheme is highly relevant. Consequently, incident reports should provide appropriate confidentiality and integrity guarantees and, ideally, be tamper-resistant or tamper-evident.

\subsection{Legislation and Standards}
In contrast to many of open research questions discussed above, which require in-depth investigation, some implications for agentic AI incident reporting can be already identified. In particular, existing reporting schemes should be expanded to incorporate the information requirements identified in this paper.

Regarding privacy, a possible solution based on research results is to adopt a fine-grained scheme depending on whether a report is publicly released, required for reproducibility or in a court case.
In general, standardized reports should be privacy-friendly by avoiding narrative text,
while providing categoric entries which remain coarse enough that each category is populated by a sufficient
number of reports.
This creates a deliberate granularity trade-off. Categories must be fine-grained enough to capture relevant distinctions between incidents, but not so fine-grained that rare category combinations become effectively identifying or disappear under the noise, thresholding, or suppression used by a
differentially private release mechanism~\cite{dwork2014algorithmic,aumueller2021sparse}. An example of such a framework is VERIS~\cite{veris2026framework}.

In general, an agreement on the right level of protection is necessary. For example, concerns about exposing IP
can lead to under-reporting, as witnessed in the Biological Weapons Convention verification gap~\cite{lentzos2019compliance}.
A counterexample is aviation occurrence reporting under Regulation (EU) No 376/2014, which pairs mandatory and voluntary schemes with just culture and mandatory de-identification, showing that a reporting duty and non-punitive treatment of the reporter are compatible.
A similar approach could support the AI Act Article 73 reporting obligations~\cite{EUAIAct2024} falling on
providers. That obligation is triggered by harm as defined in Article 3(49), not by compromise, so the agentic vulnerabilities cited above fall outside it unless they produce such harm.
\section{Conclusion} 
This paper synthesizes the perspectives of \textbf{23} academic and industry experts on AI agents security incident reporting. We have identified open research questions concerning the efficiency and effectiveness of agentic AI incident reporting, as well as the information needed to support the detection of incidents in agentic AI systems. We also derived a preliminary reporting scheme (see Table~\ref{tab:overviewElements}), that specifies the information to be captured in such reports. 
Expert input further highlighted general security and confidentiality challenges that may arise when reporting incidents involving AI agents, including information leakage and adversarial attacks targeting the reporting process itself.
We have summarized the implications of these findings for legislation and standardization, highlighting both the changes needed in existing frameworks and the challenges the research community must address to enable trustworthy AI incident reporting and thus trustworthy AI. Without reporting mechanisms that capture autonomy, delegation, memory, and agent interactions, many agentic AI security incidents will remain difficult to understand, reproduce, and ultimately prevent.

\section{Acknowledgments}
The authors would like to thank Rudolf Mayer to connect Kathrin and Anastasia for this project. Juliane Hoffmann acknowledges support by Deutsche Forschungsgemeinschaft (DFG, German Research Foundation) as part of
the Research and Training Group 2475 "Cybercrime and Forensic Computing" (grant number 393541319/GRK2475/2-2024).

\clearpage
\begin{table}[t]
\centering
\caption{Agentic AI System Elements Relevant to Security Incident Reporting.}
\label{tab:overviewElements}
\begin{tabular}{p{0.0cm}p{6cm}p{8cm}}
\toprule
& \textbf{Element} & \textbf{Description} \\
\midrule

\multicolumn{3}{l}{\textbf{Describing the agent and its trajectory}} \\
& Message histories & Full record of messages between agent and user \\ 
& Reasoning tokens & All tokens produced during reasoning steps \\
& Data provenance & Full provenance and/or version of external data, or data itself if custom  \\
& Data writes/reads & All data read and all changes written \\
& Theoretical autonomy & Possible autonomy allowed by design \\
& Effective autonomy & Actual autonomy exercised at time of incident \\ 
\addlinespace
\multicolumn{3}{l}{\textbf{Describing Agent's Capabilities}}  \\
& Defined Trust boundaries & Trust environments, limits in high-trust elements, etc. \\ 
& Available tools & Accessible tools by design, with versions \\
& Effective tool use & Exact tools invoked, arguments passed, responses received \\
& Interaction with OS & Any actions taken on the underlying operating system \\ 
& Agent identity & For permissions, tool use, etc \\
& Delegation chain & Full chain of delegation from human/agent initiator \\
& Authentication tokens & Tokens used for authentication or tool use \\ 
\addlinespace
\multicolumn{3}{l}{\textbf{Agent sub-components and orchestration}} \\
& Defined Multi-agent architecture & Defined architecture with all agents and their function \\
& Actual Multi-agent architecture & Invoked agents and their function during trajectory and incident \\ 
& Orchestration protocols & Orchestration protocols and coordination mechanisms in use \\
& Communication logs & All inter-agent messages exchanged during execution \\ 

\bottomrule
\end{tabular}

\end{table}
\clearpage

\bibliography{lit}

\end{document}